\documentclass[preprint,aps,prd, showpacs, amsmath,amssymb,eqsecnum]{revtex4}
\usepackage{graphics}
\usepackage{graphicx}%
\usepackage{wrapfig}
\usepackage{subfigure}
\usepackage{mathptmx}
\usepackage{psfig}
\usepackage{dcolumn}
\usepackage{bm}
\usepackage{tabls}
\newcommand{\be}{\begin{eqnarray}}
\newcommand{\ee}{\end{eqnarray}}
\long\def\hidestart#1\hideend{}
\begin{document}
\title{Topological charge in 1+1 dimensional lattice $\phi^4$ theory}

\author{Asit K. De}
\email{asitk.de@saha.ac.in}

\author{A. Harindranath}
\email{a.harindranath@saha.ac.in}

\author{Jyotirmoy Maiti}
\email{jyotirmoy.maiti@saha.ac.in}

\author{Tilak Sinha}
\email{tilak.sinha@saha.ac.in}

\affiliation{Theory Group, Saha Institute of Nuclear Physics \\
 1/AF Bidhan Nagar, Kolkata 700064, India}

\date {October 6, 2005}

\begin{abstract}

We investigate the topological charge in 1+1 dimensional $\phi^4$
theory on a lattice with Anti Periodic Boundary Condition (APBC)
in the spatial direction. We
propose a simple order parameter for the lattice theory
with APBC and we demonstrate its effectiveness. Our study suggests that kink
condensation is a possible mechanism for the order-disorder phase
transition in the 1+1 dimensional $\phi^4$ theory. With renormalizations
performed on the lattice with Periodic Boundary Condition (PBC), the
topological charge in the renormalized theory is given as the ratio of
the order parameters in the lattices with APBC and PBC. 
We present a comparison of topological charges in the bare and the
renormalized theory and demonstrate invariance of the charge of the 
renormalized theory in the broken symmetry phase.

\end{abstract}

\pacs{02.70.Uu, 11.10.Gh, 11.10.Kk, 11.15.Ha}
\maketitle


\section{Introduction}
Topology has an important role in nonperturbative quantum field theory on
a lattice. For early work see, for example, \cite{earlyref}. It is of
utmost importance in the dual super-conductivity picture of quark
confinement in QCD \cite{digiacomo}. In field theories with non-trivial
topology, different topological sectors are characterized by a
conserved topological charge which depends on the boundary behaviour 
of the theory. This conserved quantity  
is well-defined in classical continuum field theory.

In 1+1 dimensional classical $\phi^4$ field theory, there exists a
kink solution with a conserved topological charge \cite{raja}. In
the quantum version of the above theory using lattice regularization,
past studies on kinks focussed on the
calculation of the kink mass \cite{ct,aw}. Recently these kinks have
also been studied in discrete light cone quantization \cite{dlcq}.  

In this paper we investigate the topological charge in the
context of the ordered (broken symmetry) phase of 1+1 dimensional 
lattice $ \phi^4$ field theory. 
The short range quantum fluctuations lead to the renormalization of 
the mass and the coupling constant in the
topologically trivial sector of the theory. On the other hand, long range 
fluctuations lead to a phase transition in the strong coupling region which
leads to a breaking of the conservation of topological current and hence the
vanishing of the topological charge in the disordered (symmetric) phase. 
We show that the topological 
charge in the renormalized theory remains invariant in the ordered
phase. To the best of our knowledge, topologocal charge and its
invariance in this
quantum theory has not been investigated before.

Given that the $\phi^4$ theory and the Ising model are in the same
universality class, one can also define a topolgical charge in the
Ising model when considered as a Euclidean quantum field theory in
1+1 dimensions \cite{digiacomo2}. However, our aim in this work is to
look at the
renormalizations of the parameters of the $\phi^4$ theory so that a
sensible definition of topological charge, invariant in the ordered
phase of the theory, emerges. 
 
Condensation of the topological objects has been associated with the mechanism
of phase transitions in various statistical and quantum field theories
\cite{digiacomo2,sf,kogut} (also see works cited in \cite{digiacomo2}). 
Our study indicates that condensation of kink-antikinks is a possible
mechanism for the order-disorder phase transition in 1+1 dimensional
$\phi^4$ theory.

The plan of this paper is as follows. In Sec. II we present our notation
and definition of the topological charge in the bare and renormalized
lattice theory. Secs. III, IV and V are devoted to the numerical work with
details of various issues that we encounter in the
calculation. Finally, Sec. VI contains discussion of our results and
conclusion. 
   
\section{Topological charge and renormalization}

We start from the Lagrangian density in Minkowski space (in usual notation)
\be 
{\cal L} = \frac{1}{2} \partial_\mu \phi \partial^\mu \phi - \frac{1}{2}
m^{2} \phi^2 - \frac{\lambda}{4!} \phi^4
\ee
which leads to the Lagrangian density in Euclidean space
\be
{\cal L}_E = \frac{1}{2} \partial_\mu \phi \partial_\mu \phi +
\frac{1}{2}
m^2 \phi^2 + \frac{\lambda}{4!} \phi^4 .
\ee

\noindent Note that in one space and one time dimensions, 
the field $\phi$ is dimensionless and
the quartic coupling $\lambda$ has dimension of mass$^{2}$. 

\noindent The Euclidean action is
\be
S_E = \int d^2x {\cal L}_E.
\ee
Next we put the system on a lattice of spacing $a$ with
\be
\int d^2x = a^2 ~ \sum_x.
\ee

\noindent Because of the periodicity of the lattice sites in a toroidal 
lattice the 
surface terms will cancel among themselves (irrespective of the boundary
conditions on fields) enabling us to write
\be 
(\partial_\mu \phi)^2 = - \phi \partial_\mu^2 \phi
\ee
and on the lattice
\be \partial_\mu^2 \phi = \frac{1}{a^2} \left [ \phi_{x+ \mu}+ \phi_{x -
\mu}
- 2 \phi_x \right ]~.
\ee
Introducing dimensionless lattice parameters $m_0^2$ and $ \lambda_0$ by
$ m_0^2 = m^2 ~ a^2 $ and $ \lambda_0 = \lambda~a^2 $
 we arrive at the lattice action in two Euclidean dimensions
\be
S = -\sum_x \sum_\mu \phi_x \phi_{x+ \mu} ~+~ (2 + \frac {m_0^2}{2})
~\sum_x ~\phi_x^2 + \frac {\lambda_0}{4!}~\sum_x ~ \phi_x^4~.
\ee 

With antiperiodic boundary condition (APBC), we need to identify an order
parameter to characterize different phases of the theory since
$ \langle \phi \rangle$ is zero in both symmetric and broken phases. 
In previous works on kinks in two dimensional
lattice field theory \cite{ct,aw}, vanishing of the kink mass in the
symmetric phase was used to
identify the critical coupling.  Since the calculation of the 
kink mass is rather
involved, it is preferable to have a simpler choice. 

For a $L^2$ lattice, we propose $ {\overline \phi_{\rm diff}} = \frac{1}{2} 
\left [ ~{\overline \phi}_{L-1} - \overline{\phi}_0 ~\right ] $ as the 
order parameter where 
$\overline{\phi}_x ~\equiv~ \langle {~\rm kink ~g.s~} | 
\phi_x | {~\rm kink ~g.s~} \rangle$
with $ | {~\rm kink ~g.s~} \rangle$ denoting the kink ground state.
$\overline{\phi}_x$ is computed by taking the average of $\phi_x$ over
configurations with APBC in the spatial direction (which effectively
performs importance sampling around the kink configurations in a
Monte-Carlo simulation).

In the classical theory, the topological charge is given by
\be
Q_{\rm classical} = \sqrt{\frac{\lambda}{6 m^2}} ~ \phi_{\rm diff} = 
\sqrt{\frac{\lambda}{3 m_B^2}} ~ \phi_{\rm diff} 
\ee
where $
\phi_{\rm diff} = \frac{1}{2} \left [ \phi(\infty) - \phi(-\infty) \right ] $
and we have used the fact that the mass of the elementary 
boson is $ m_{B} = \sqrt{2} m $ in the broken phase.
In the classical theory, the relation between $ \phi_{\rm diff}, ~m_B$ and $
\lambda$ is given by 
\be
 \phi_{\rm diff} = \sqrt{\frac{3 m_{B}^2}{\lambda}} \label{phi} 
\ee
which guarantees that the topological charge is
+1 (kink sector) and -1 (antikink sector) in the broken symmetric phase 
and zero in the symmetric phase.

In the lattice theory with APBC in the $``$spatial" direction, 
the bare topological charge is defined by 
\be Q_0 = \sqrt{\frac{\lambda_{0}}{3 m_{B_{0}}^2}} ~ {\overline \phi}_{\rm
  diff}.\label{qb}  
\ee

\noindent Due to quantum fluctuations, the classical relation 
analogous to Eq. (\ref{phi}) 
between $ {\overline \phi}_{\rm diff}, ~m_{B_0} $ and $\lambda_0$ is not
obeyed and as a consequence,  $Q_0$ does not remain +1 or -1 in the ordered
phase.

We propose to define the topological charge in the renormalized theory as
\be Q_{R} = \sqrt{\frac{\lambda_{R}}{3 m_{R}^2}} ~ {\overline \phi}_{R_{\rm
diff}} \label{qr}
\ee 
where $ \phi_R = \frac{1}{\sqrt{Z}} \phi $, $Z$ being the field 
renormalization constant.
The conventional definition of the renormalized coupling $\lambda_R$ in 
$\phi^4$ theory 
is in terms of  
the renormalized four-point vertex function. The calculation
of the four-point vertex function on the lattice is computationally
demanding. Fortunately,
in the broken phase, we can choose a definition \cite{munster} 
of $ \lambda_{R}$
in terms of the renormalized mass $m_R$ and the renormalized vacuum 
expectation value 
$ \langle \phi_{R} \rangle $, determined with 
periodic boundary condition (PBC):
\be
\lambda_{R} = 3 \frac{m_R^2 }{\langle \phi_R \rangle^2} .
\ee
This definition of $ \lambda_R$ involves only the computation of 
one-point function for $ \langle \phi \rangle $
and the two point function for $m_R$ and $Z$. The details of these
computations are provided in 
Ref. \cite{longpaper}. 
Using this definition of $ \lambda_{R}$ in Eq. (\ref{qr}), we get
\be
 Q_{R} =  \frac{{\overline \phi_{R_{\rm diff}}}}{\langle \phi_{R} \rangle } =
\frac{{\overline \phi_{\rm diff}}}
{\langle \phi \rangle}. \label{qr1}
\ee 
The second equality of Eq. (\ref{qr1}) assumes $Z$ to be the same for
$\phi_{\rm diff}$ and $\phi$ and therefore involves two unrenormalized 
quantities, $\overline \phi_{\rm diff}$ 
calculated with APBC and $\langle \phi \rangle$ calculated with PBC. If 
$\overline \phi_{\rm diff}$ and $\langle \phi \rangle$ are of the same
magnitude, the renormalized   topological charge is $ \pm 1$ in the ordered
phase except for the  statistical and systematic errors that occur in the
numerical computations of the quantities in the numerator and the
denominator of Eq. (\ref{qr1}).  
\section{Calculation with PBC: $\langle \phi \rangle$}
\begin{figure}
\begin{minipage}[t]{8cm}
\includegraphics[width=1\textwidth]{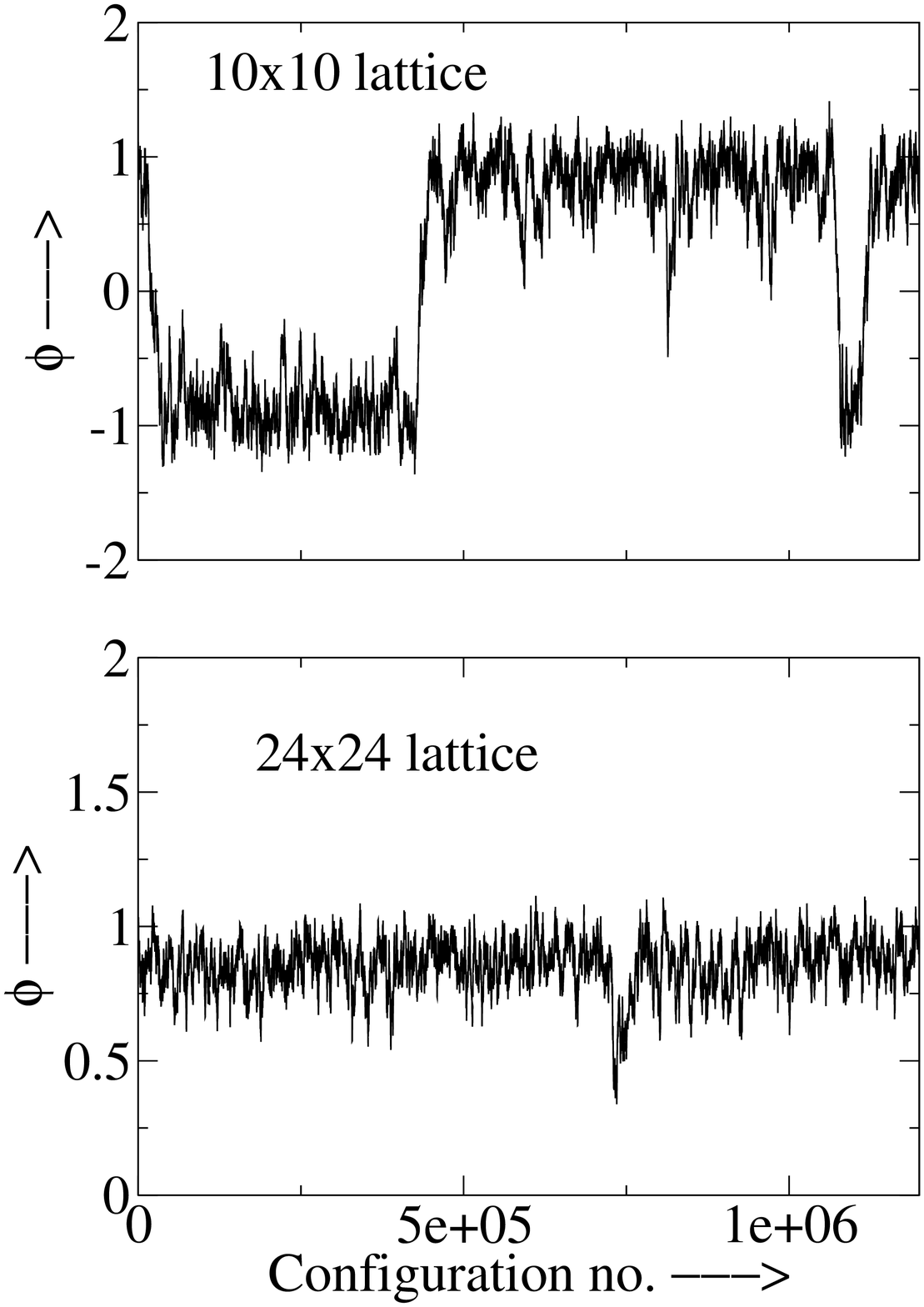}
\caption{Comparison of tunneling in $\phi^4$ theory for couplings
  $m_0^2= -0.5,~\lambda_0= 1.8$.}             
\label{tunnel}
\end{minipage}
\hfill
\begin{minipage}[t]{8cm}
\includegraphics[width=1\textwidth]{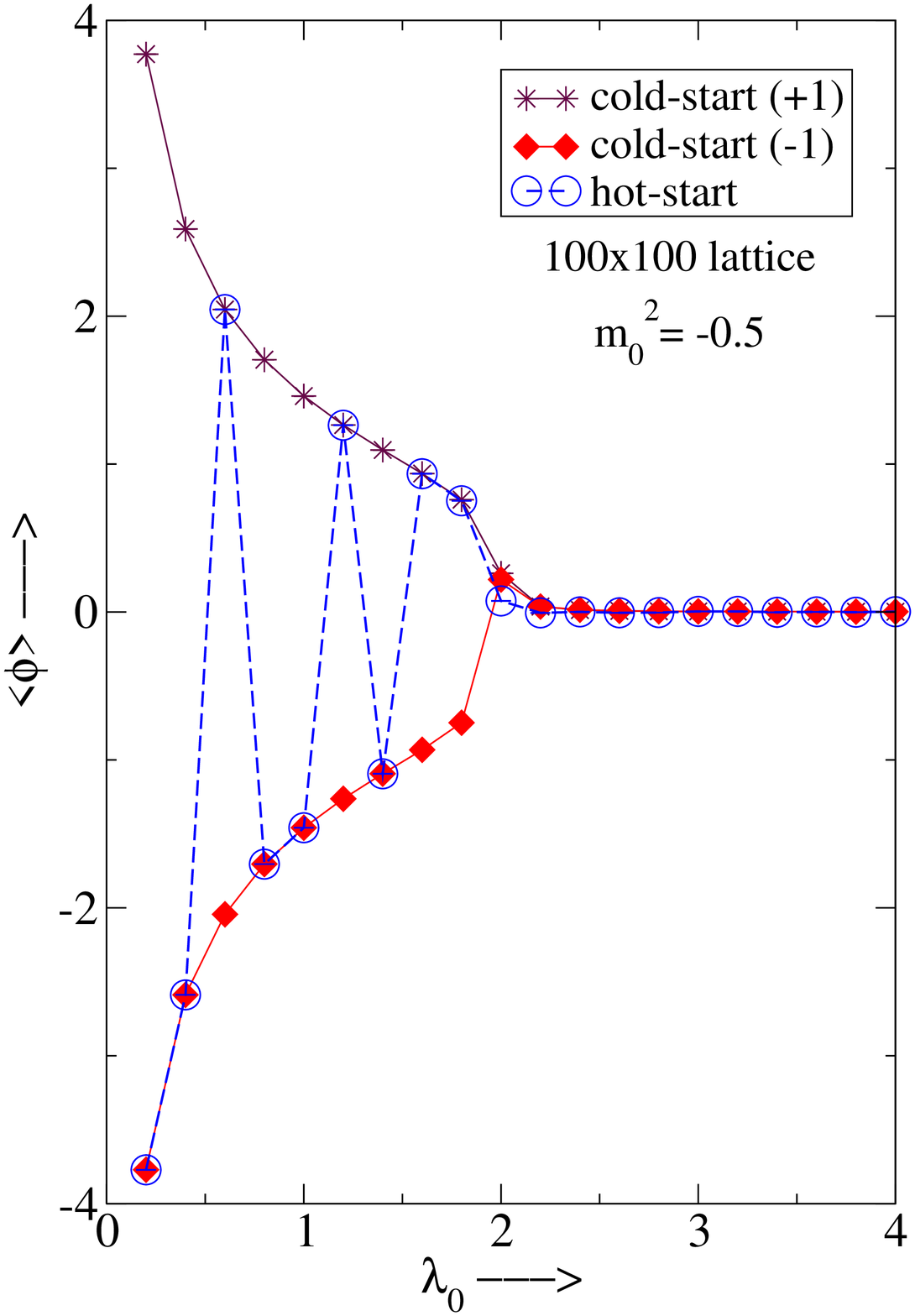}
\caption{Sensitivity to initial configuration (Metropolis algorithm).}        
\label{diff-start}
\end{minipage}
\end{figure}

First we discuss the calculation of $\langle \phi \rangle$ with PBC. 
As is well-known, spontaneous symmetry breaking cannot occur at finite
volume due to the quantum mechanical tunneling phenomena. This is due to
the fact that at finite volume we have  finite degrees of freedom.
Tunneling probability is, however, inversely proportional to the 
exponential of the volume and hence is exponentially suppressed at large
volume \cite{munster,montvay}. Thus in
practice, the issue is how large a volume we need for the suppression to
occur. 

For $ \phi^4$ theory, tunneling is present for $10^2$ lattice but is
practically absent for $24^2$ lattice. This is demonstrated in 
Fig. \ref{tunnel} where $\phi$ averaged over all sites for a given
configuration is plotted against
the configuration number. These calculations were done using single hit
Metropolis algorithm.    

We find $\langle \phi \rangle$ to be sensitive to the various choices
of the initial configuration (see Fig. \ref{diff-start}) in the simulation 
using the Metropolis algorithm. For cold
starts [$\phi$ = +1 (or -1) for all sites] the selection of the vacuum is
controlled by the choice of the sign of $\phi$ irrespective of the 
couplings. On the other hand for hot starts ($\phi=\pm1 
$ sprinkled randomly over the entire lattice), 
the vaccum is picked randomly from the two degenerate vacua.
\begin{figure}
\begin{minipage}[t]{8cm}
\includegraphics[width=1\textwidth]{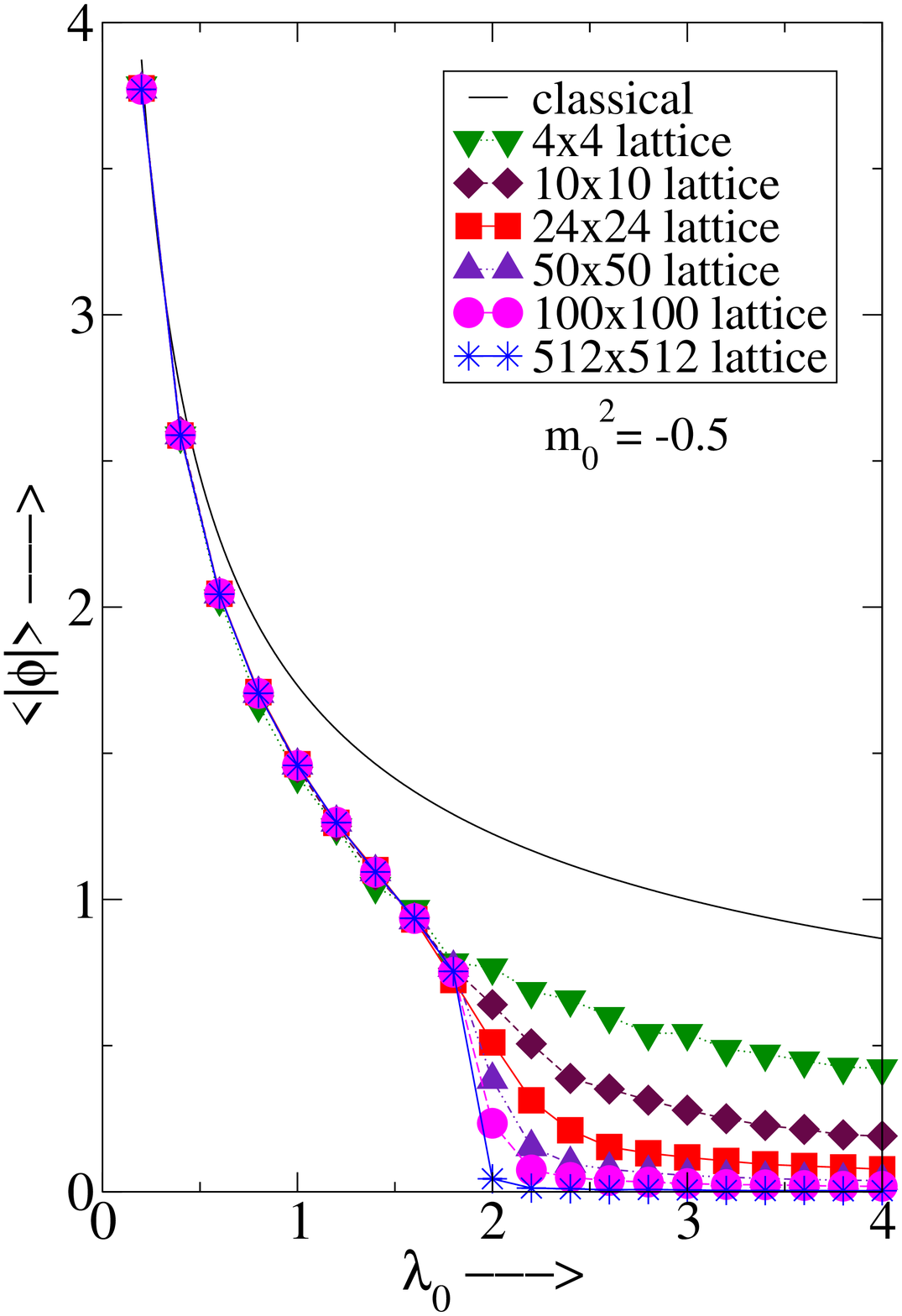}
\caption{$\langle |\phi| \rangle$ versus $\lambda_0$ for various
lattice volumes. For comparison, classical result is also given.}
\label{diff-vol}
\end{minipage}
\hfill
\begin{minipage}[t]{8cm}
\includegraphics[width=1\textwidth]{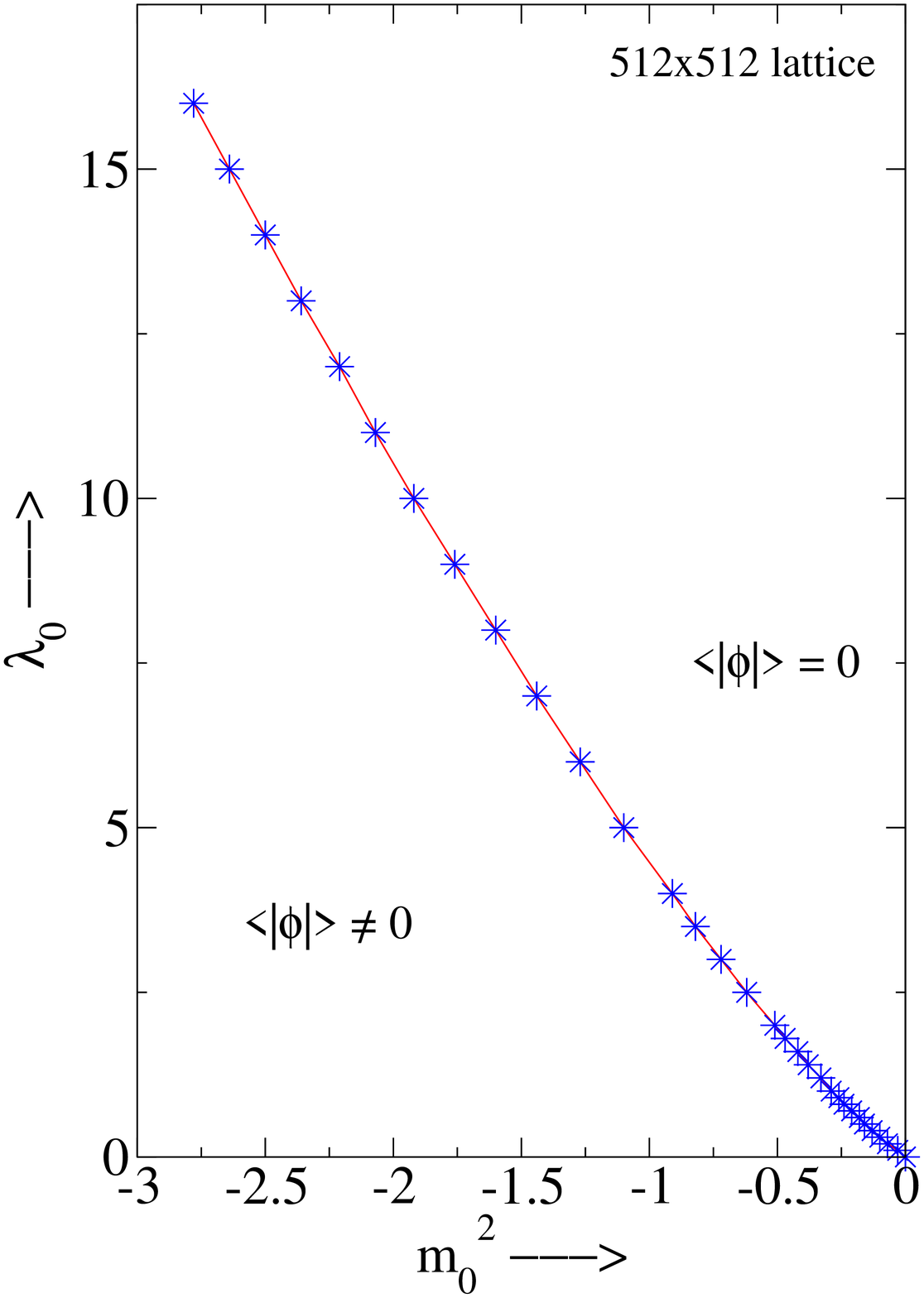}
\caption{Phase diagram.}
\label{phase}
\end{minipage}
\end{figure}

A more serious problem with the Metropolis algorithm is critical slowing
down which makes computation near the critical point 
prohibitively expensive. In the
Monte-Carlo simulation, to reduce critical slowing down, we have thus 
resorted to the standard
Metropolis update {\em combined} with a cluster algorithm \cite{wolff} 
to update the embedded Ising variables in $\phi^4$ theory, following
the method of Brower and Tamayo \cite{bt}. We use Wolff's single cluster
variant of the algorithm. This method has been used
previously in determining the phase diagram of two dimensional lattice
$\phi^4$ theory \cite{lw}.  

In the cluster algorithm however, since the
sign of the field of all the members of the cluster are flipped in every
updation cycle, the algorithm actually enforces tunneling
and the configuration average of $ \phi$ is always zero. Thus
to get the appropriate nonzero value for the condensate we measure 
$\langle |\phi| \rangle$ where 
$\phi = \frac{1}{\rm Volume}\sum\limits_{\rm sites}\phi(x)$. To understand the
mod let us 
consider a local order parameter $\langle  \phi(x) \rangle$. Since
the configurations will be selected at random dominantly 
from the neighborhood of
either vacua in the broken phase, $\langle \phi(x) \rangle$ 
will vanish when averaged
over configurations thus wiping out the signature of a broken phase. If one
uses $\langle |\phi(x)| \rangle$ as the order parameter then in the
broken phase it correctly projects itself onto one of the vacua yielding
the appropriate non-zero value. The use of this mod, unfortunately, destroys
the signal in the symmetric phase completely by wiping out the significant
fluctuations in sign. However if we choose to use 
$\langle |\frac{1}{\rm Volume}\sum\limits_{\rm sites}\phi(x)| \rangle$, it 
correctly captures the broken phase as well as the symmetric phase. 
While the sign fluctuation over configurations are still masked, the 
fluctuations over sites survive producing 
$\langle |\phi| \rangle = 0$ correctly in
the symmetric phase.
     
The volume dependence of $\langle |\phi| \rangle$ is presented in
Fig.\ \ref{diff-vol} where we also compare with the classical
value of the condensate. As can be seen, for small volumes, the signal for
phase transition is very weak to detect, but $50^2$ lattice is big enough to
observe the transition.   

In Figs. \ref{diff-start} and \ref{diff-vol} and all the figures to follow
the standard error bars, if not visible, are smaller than the symbols
unless otherwise stated.

The phase diagram we obtained for a $512^2$ lattice using PBC
is presented in Fig. \ref{phase}. This agrees with the phase diagram
obtained in \cite{ct,aw,lw}. In \cite{lw} the authors extrapolate their results
to infinite volume. We observe that our $512^2$ lattice results are 
as good as the infinite volume result in Ref. \cite{lw}. 

Most of our calculations are performed at $m_0^2= -0.5$. The
corresponding critical coupling $\lambda_0^c$ is around 1.95. 
\section{Calculation with APBC: Kink configurations and ${\overline
\phi_{\rm diff}}$} 
Cluster algorithms are known to fail with antiperiodic boundary conditions
in $\phi^4$ theory \cite{hasen}.
For the calculation of ${\overline \phi_{\rm diff}}$, because of APBC, 
we cannot use the cluster algorithm and we resort to the
standard Metropolis algorithm.  In addition to the problems associated with
critical slowing down, we  also face problems associated with computing the
profile of an extended object, the topological kink in this case and also 
in fixing its location for the measurement of topological charge.

\begin{figure}
\begin{minipage}[t]{8cm}
\includegraphics[width=1\textwidth]{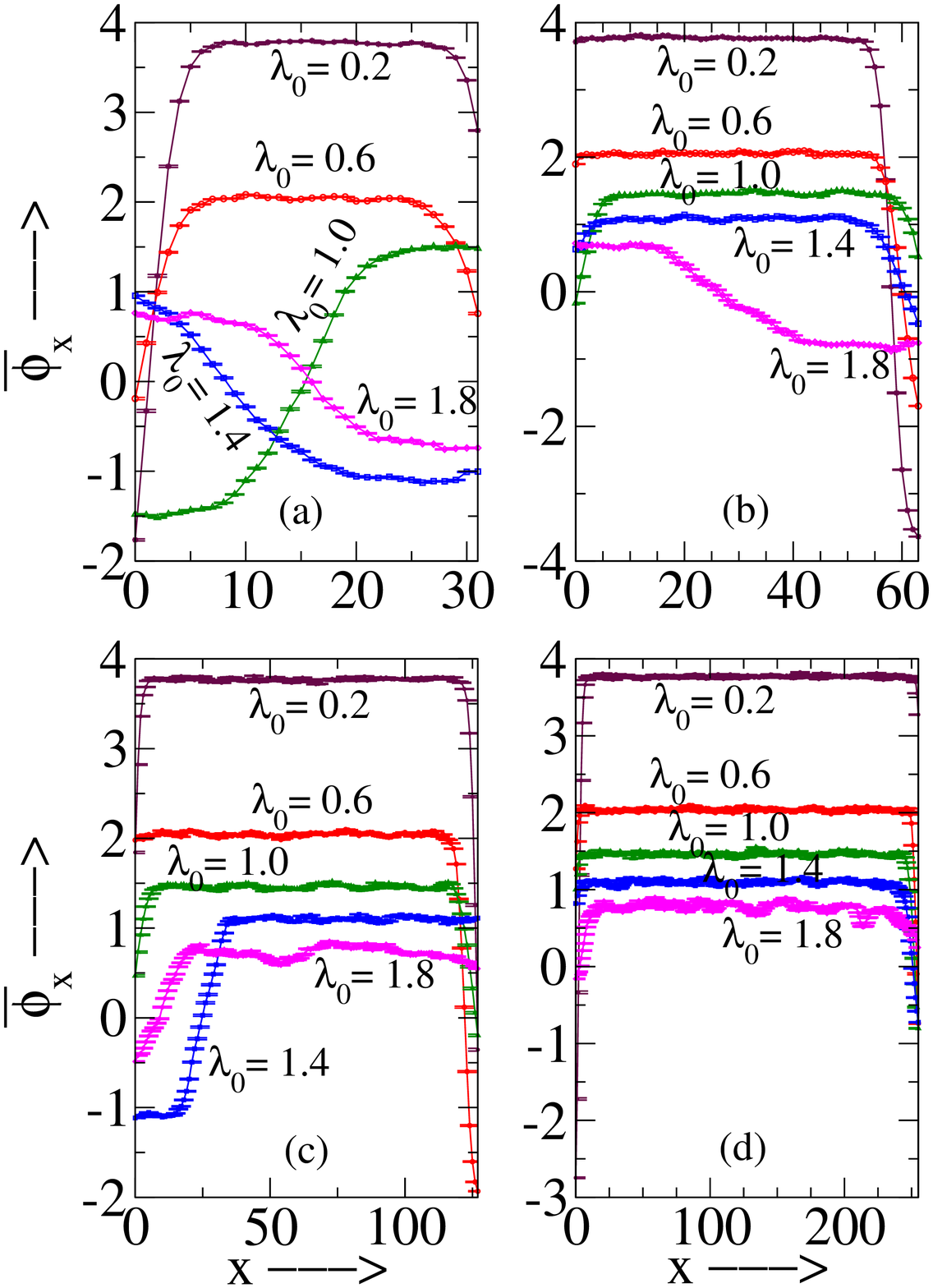}
\caption{${\overline \phi}_x$ for cold-start(+1) at $m_0^2= -0.5$ for lattices
(a) $32^2$, (b) $64^2$, (c) $128^2$ and (d) $256^2$.}
\label{cold}
\end{minipage}
\hfill
\begin{minipage}[t]{8cm}
\includegraphics[width=1\textwidth]{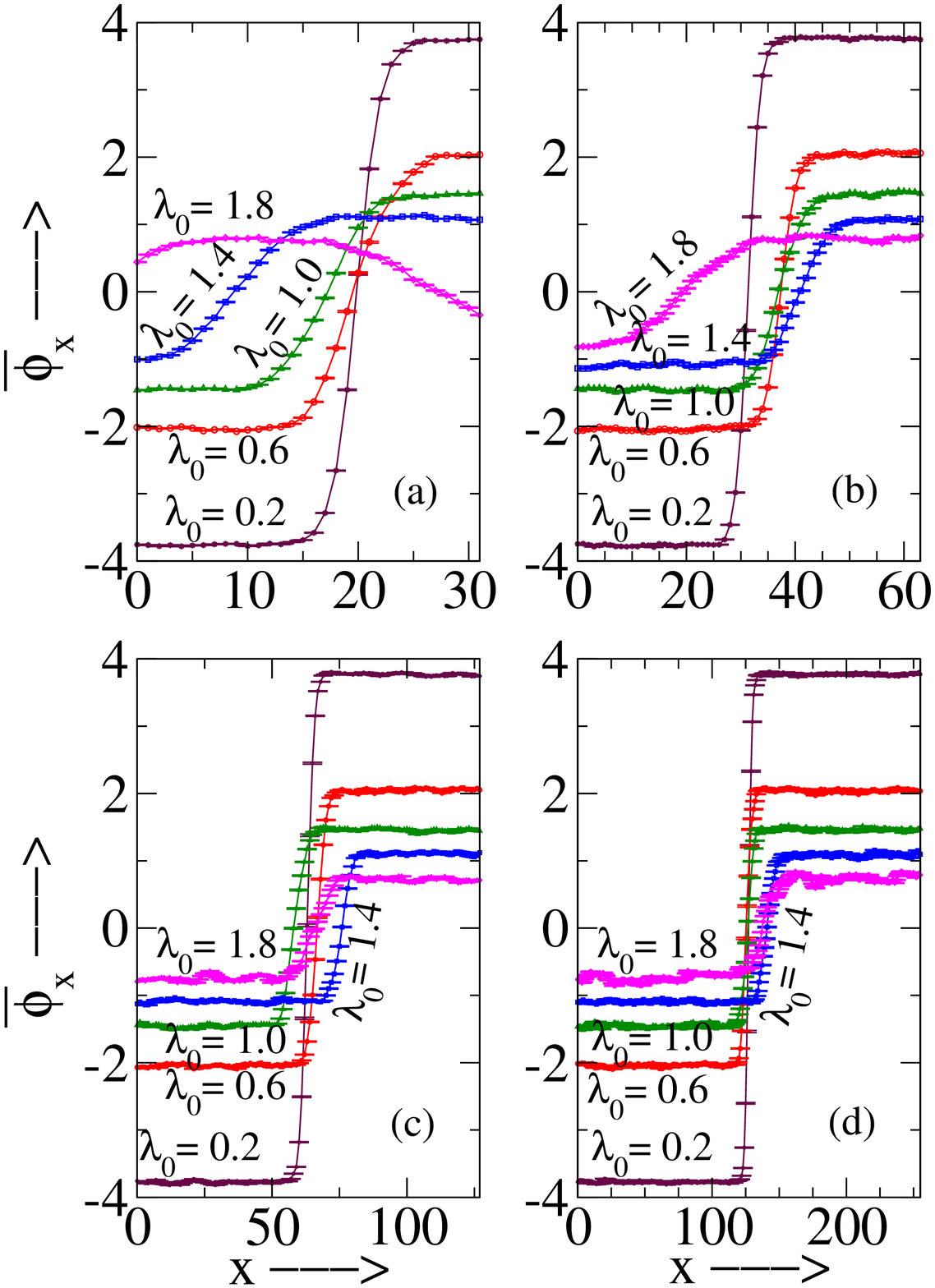}
\caption{${\overline \phi}_x$ for kink-start at $m_0^2= -0.5$ for
  lattices (a) $32^2$, (b) $64^2$, (c) $128^2$ and (d) $256^2$.} 
\label{kink}
\end{minipage}
\end{figure}

\begin{figure}
\begin{minipage}[t]{8cm}
\includegraphics[width=1\textwidth]{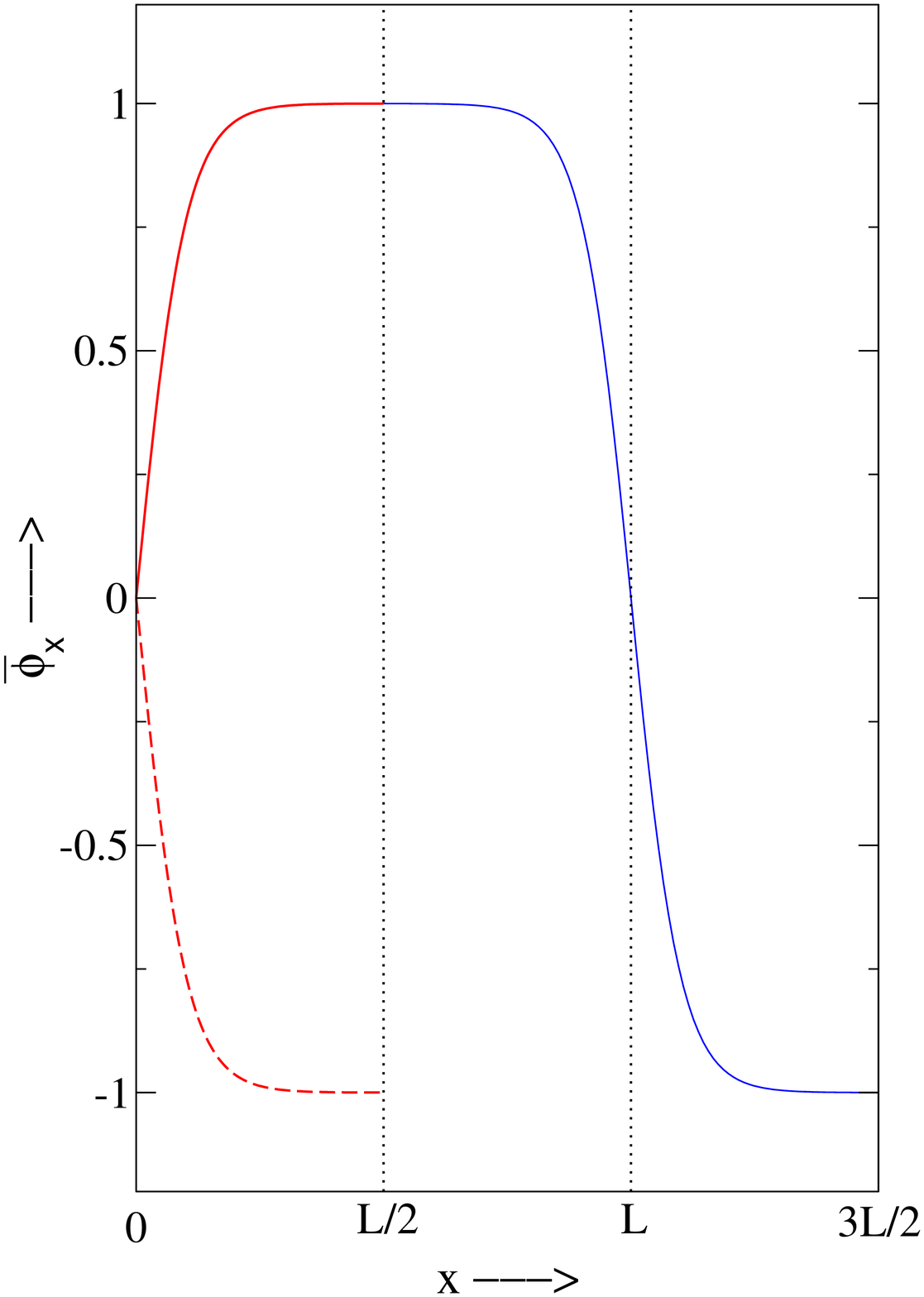}
\caption{The part of the kink between $L$ to $3L/2$ 
gets translated to $0$ to $L/2$ (short dash) due to periodicity of toroidal 
lattice and finally gets inverted (solid line between $0$ to $L/2$)
due to APBC on the field.}
\label{scheme}
\end{minipage}
\hfill
\begin{minipage}[t]{8cm}
\includegraphics[width=1\textwidth]{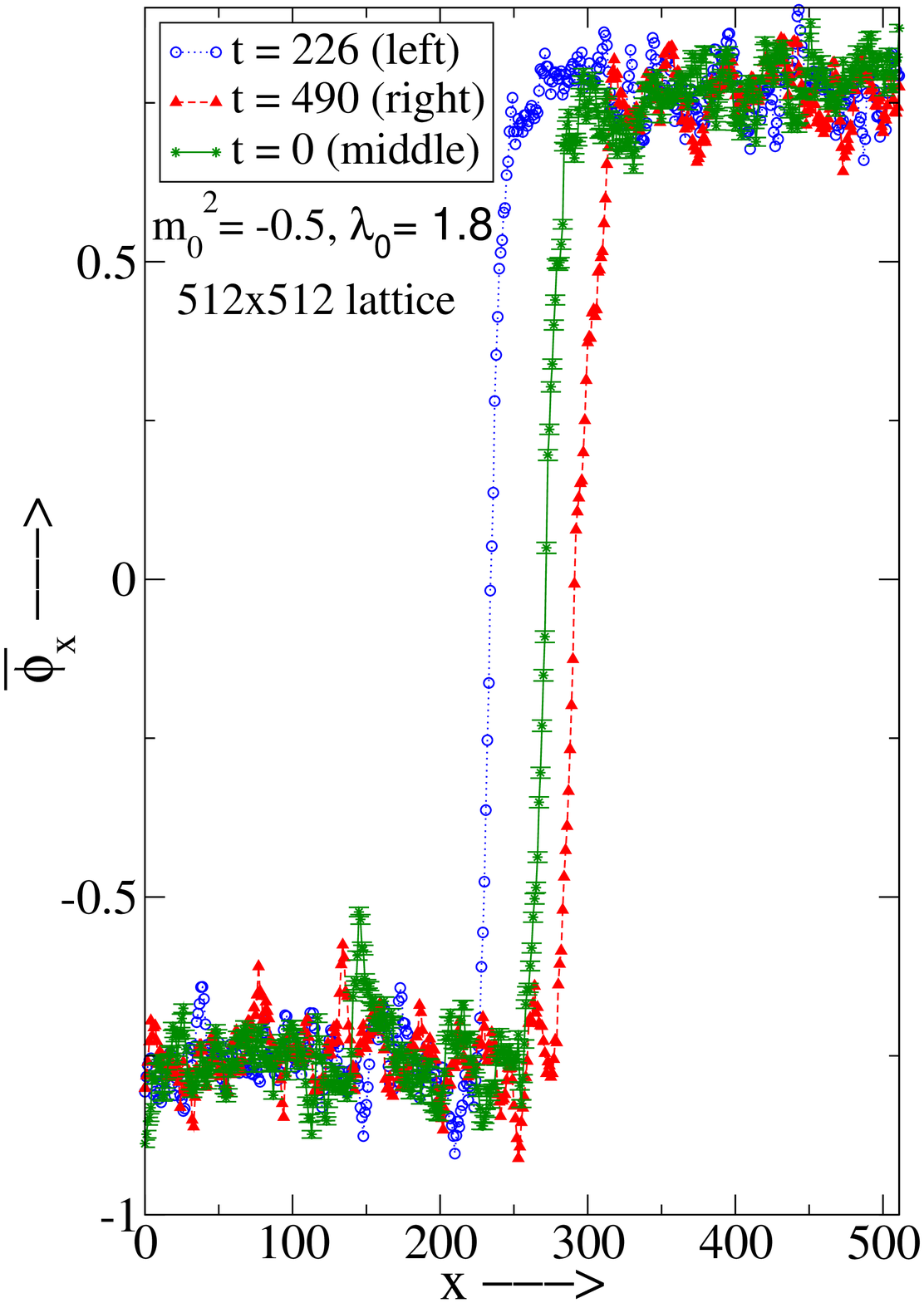}
\caption{${\overline \phi}_x$ for kink start for three different time
slices showing motion of the kink.}
\label{kinkmove}
\end{minipage}
\end{figure}

First we note that the location of the kink cannot be controlled 
with a cold start as we show in Fig. \ref{cold}. It is perhaps not easy 
at first sight, to recognize that most of the configurations 
seen in Fig. \ref{cold} are actually  kink configurations. 
To bring it out  we 
refer to Fig. \ref{scheme} in which we demonstrate schematically, how a kink 
centered near the boundary of a toroidal lattice with APBC on the fields
should look like. Note that the $(L+i)$ th site is identified with the 
$i$ th site
on a toroidal lattice but the sign of the field is flipped on account of 
APBC on the fields giving it a plateau-like
appearance. Most of the configurations in Fig. \ref{cold} resemble this
plateau.  

All such profiles in Fig. \ref{cold} are therefore basically kinks or
antikinks located near the edge. We think that the formation of kinks near
the edge are favored by the algorithm because it is clearly much easier to
generate the plateau-like configurations from a cold start ($\phi=+1$)
compared to a kink which involves a flipping of the sign of fields over larger
region. Near the
critical region of course, we expect the formation of kink like
configurations to be much easier and it is consistent with our observation
($\lambda_0=1.8$ in Fig. \ref{cold}). Let us mention at this point that
because of translational invariance,
the position of a kink is of no significance as long as it does not
tend to the spatial infinities.  

To make our results more transparent we intended to work with kink
configurations that are centered near the middle of the lattice.  
The definition of the topological charge that we use (Eq. (\ref{qr})) also 
presupposes that our kink is actually located near the center.  

To obtain such kink configurations we generated configurations by 
using the kink start ($\phi_x = -1$ for $0\le x < \frac{L}{2}$ and $\phi_x =
+1$ for $\frac{L}{2}\le x < L$) which nicely
generates the kink configurations near the center (Fig. \ref{kink}).

In passing we remark that
the data in both, Fig. \ref{cold} and Fig. \ref{kink}
are seen to deteriorate with smaller volumes owing to finite size effects 
which are as usual more serious near the critical point (large coupling in this
case). We have also observed that there is a small but noticeable increase
in the {\em size} of the kinks (spatial extent over which $\phi_x$ changes) 
with an increase of $\lambda_0$ which affects the sharpness of the kinks 
in a small volume near the critical region. 

Reasonably good 
kink configurations for the measurement of topological charge (with stable 
flat regions) are obtained with the kink start with $128^2$ and $256^2$
lattices. 

\begin{figure}
\begin{minipage}[t]{8cm} 
\includegraphics[width=1\textwidth]{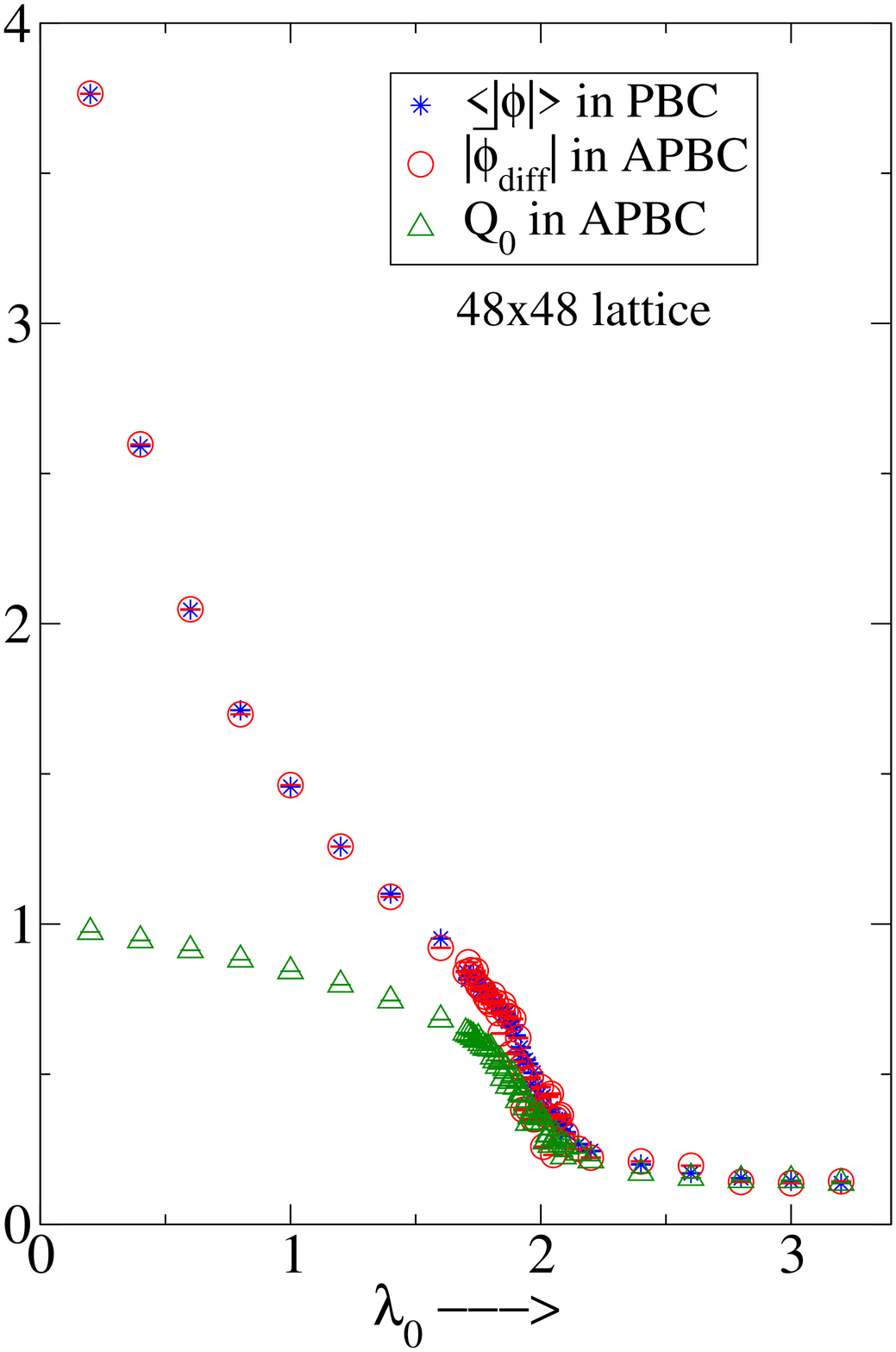}
\caption{$ \langle |\phi| \rangle$ for PBC and $|{\overline \phi}_{\rm diff}|$
for APBC versus $\lambda_0$ at $m_0^2= -0.5$. $Q_0$ for APBC also plotted.} 
\label{phicomp}
\end{minipage}
\hfill
\begin{minipage}[t]{8cm}
\includegraphics[width=1\textwidth]{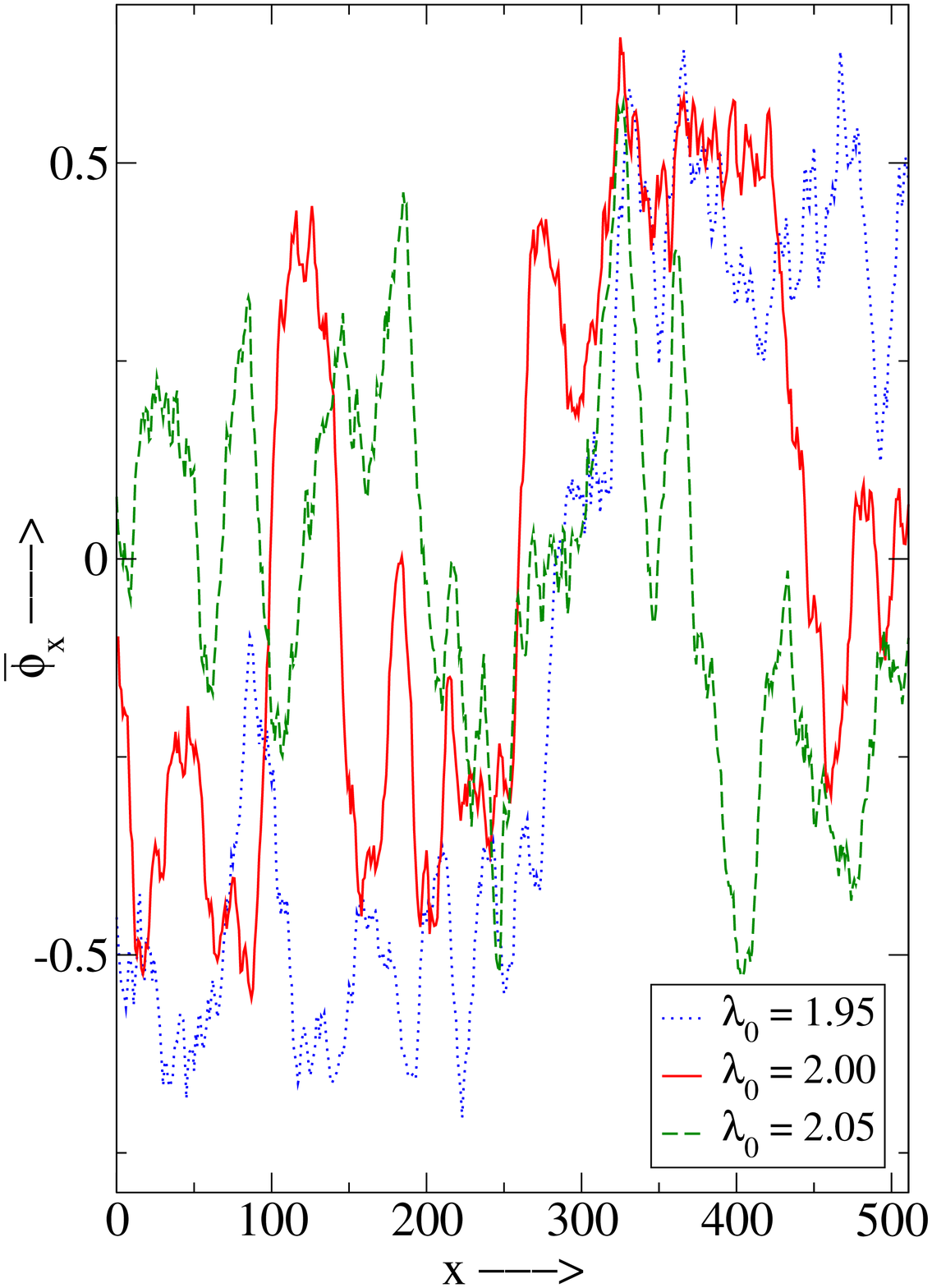}
\caption{Kink condensation shown for $512^2$ lattice at $m_0^2= -0.5$.}
\label{condense}
\end{minipage}
\end{figure}

We have observed kink motion in the vicinity of the critical region. 
This motion of the kink is clearly visible in  Fig. \ref{kinkmove} where we
present ${\overline \phi}_x$ for a $512^2$ lattice for three
different time slices. Probably because the kink mass is small
enough at the couplings, we see this motion. This is consistent with the
vanishing of the kink mass at the critical point. Incidentally, the
equality in the magnitude of ${\overline \phi}_{\rm diff}$ among different
time slices demonstrates the expected conservation of topological
charge. One would have ideally liked to increase the statistics 
by averaging over the time slices at a given $x$, but the movement of kink
makes this not viable in general. However, taking into account the fact 
that the location of the kink is of no consequence in the measurement of 
topological charge, we have actually forced the formation of kinks 
at the middle (by fixing the value of the field at the mid-point 
in every time slice, to zero) and then taken the average over time slices.   
\section{Kink condensation and topological charge}
Let us now come to the discussion of the topological charge.
For the use of Eq. (\ref{qr1}) for the topological charge, it is necessary
that we get the same phase diagram and critical behaviour for the numerator
and the denominator of this expression. 
The simulation results with PBC and APBC
shown in Fig. \ref{phicomp} for $48^2$ lattice indeed confirm this within
our numerical accuracy. Here we would also like to remark that 
Fig. \ref{phicomp} shows that the disappearance
of the kink configuration (characterized by $|\overline \phi_{\rm diff}|
= 0$) and the  
onset of the phase transition from broken to symmetric phase (characterized
by $\langle |\phi| \rangle = 0$) coincide. It therefore suggests 
`kink-condensation' as a possible mechanism for the order-disorder 
phase transition in 1+1 dimensional $\phi^4$ theory. Such a connection
is known to exist in 
2-dimensional Ising model \cite{sf,kogut,digiacomo2} which is believed to be
in the same universality class and is not totally unexpected in $\phi^4$ 
theory since it has an {\em embedded} Ising variable. Condensation of
topological excitation has generally been associated with the mechanism of
phase transition in many statistical and quantum field theories. Kink
condensation is clearly visible in Fig. \ref{condense} where
we present $ {\bar \phi}_x $ for three couplings very close to the
critical region. At $\lambda_0=1.95$,  kink configuration is just barely
visible, i.e., the boundary value
of $ {\bar \phi}_x $ is still nonzero and $ {\bar \phi}_x $ passes through
zero only once. As the coupling increases, the boundary value
of $ {\bar \phi}_x $  reduces to zero signalling disappearance of the
single kink. However, $ {\bar \phi}_x $ passes through zero
many times showing closely packed kink-antikink configurations
of varying amplitudes.

\begin{figure}
\begin{minipage}[t]{8cm}
\includegraphics[width=1\textwidth]{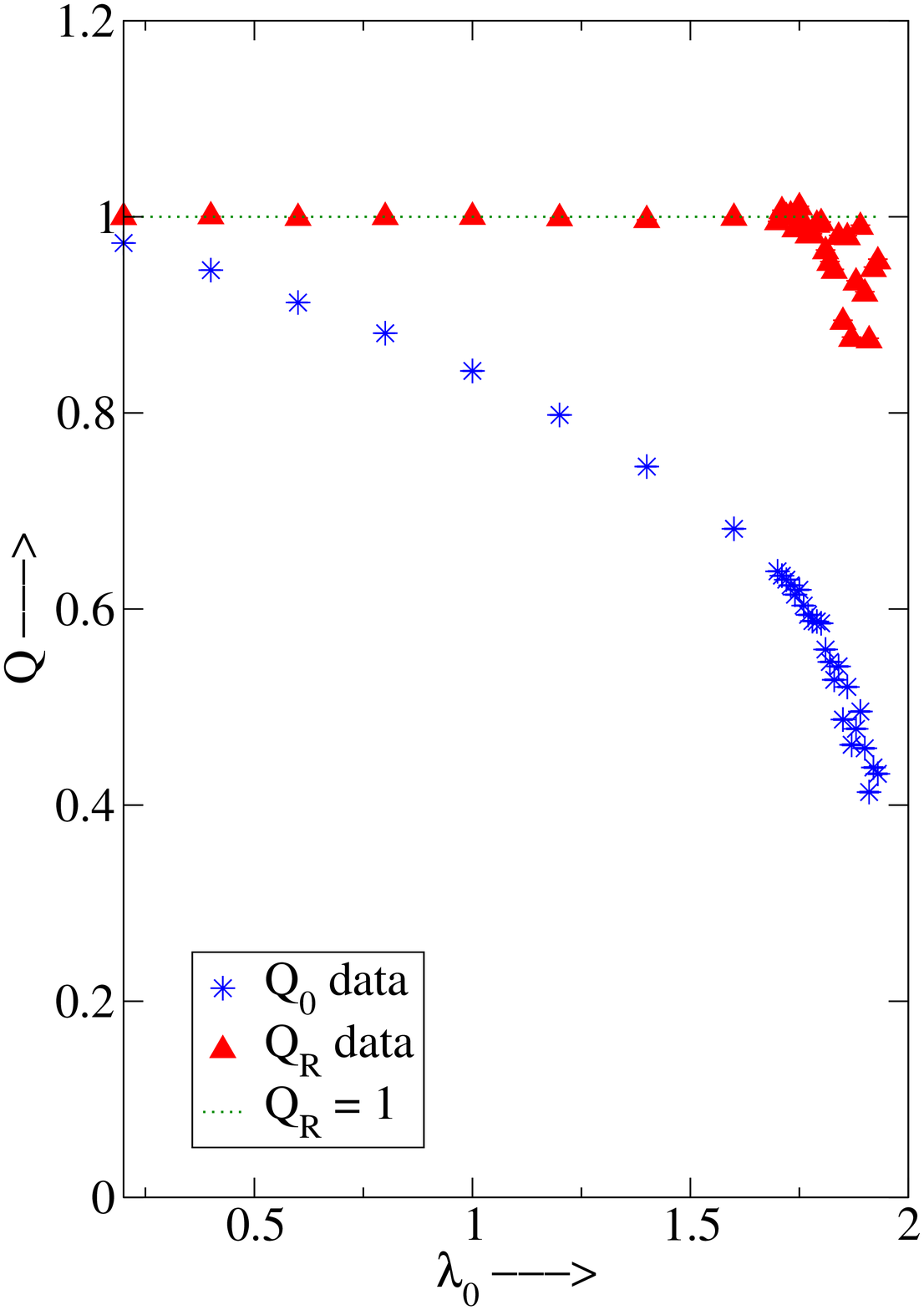}
\caption{Topological charge calculated with $48^2$ lattice at $m_0^2= -0.5$.}
\label{qcomp1}
\end{minipage}
\hfill
\begin{minipage}[t]{8cm}
\includegraphics[width=1\textwidth]{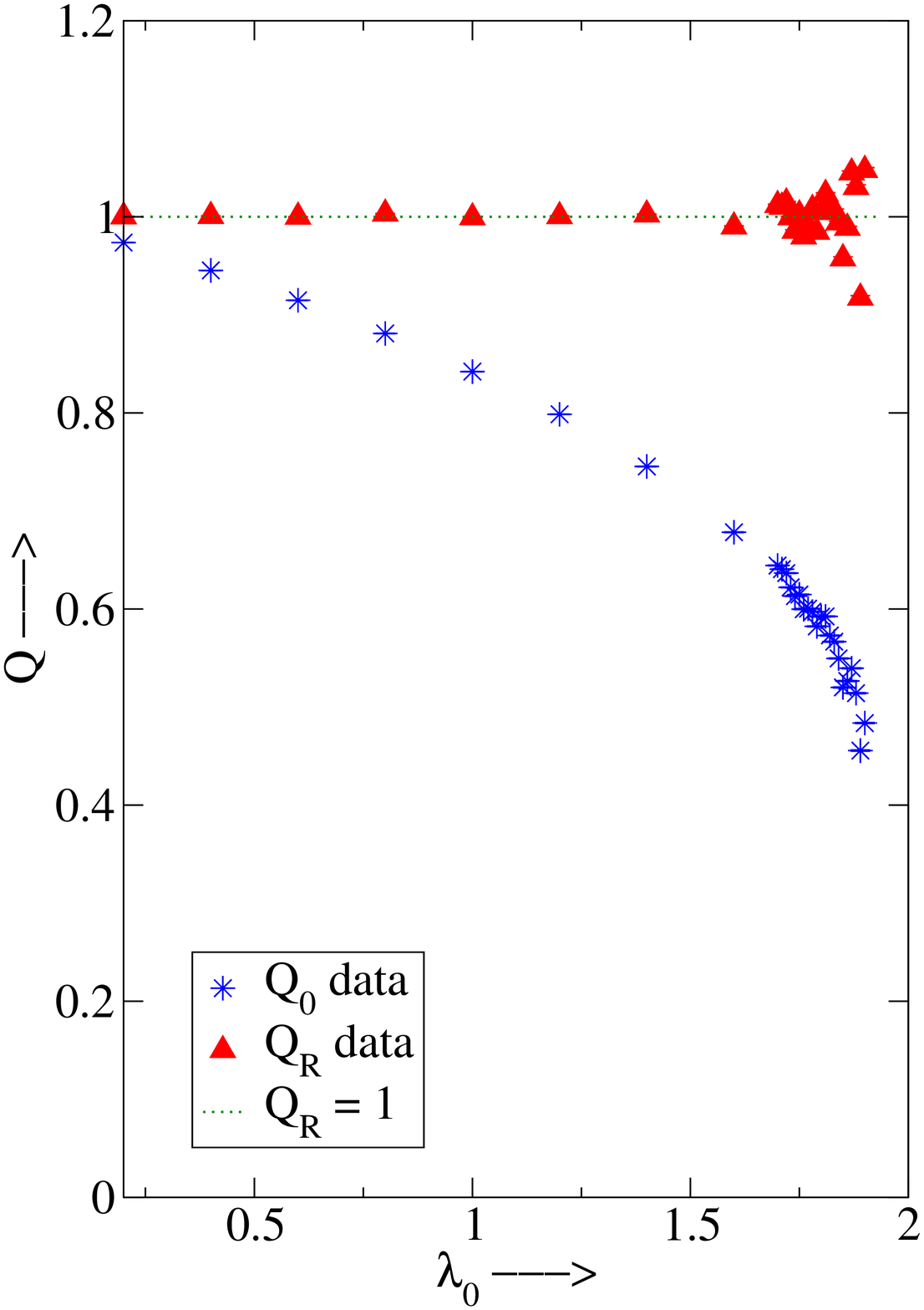}
\caption{Topological charge calculated with $512^2$ lattice at $m_0^2= -0.5$.}
\label{qcomp2}
\end{minipage}
\end{figure}

Finally we present the result for the topological charge in Figs.
\ref{qcomp1} and \ref{qcomp2}. 
The behavior of the topological charge for a range of
$\lambda_0$ is depicted in Fig. \ref{qcomp1} and Fig. \ref{qcomp2}
for $48^2$ and $512^2$ lattices respectively. The figures clearly
demonstrate the need for renormalization: It restricts $Q_R$ to +1
for the whole range of $\lambda_0$ investigated in conformity with our
expectations. Near the critical region the fluctuation of $Q_R$
around unity is due to different systematic and
statistical errors associated with different numerical algorithms used for the
calculation of the numerator and the denominator of Eq. (\ref{qr1})
especially because the numerator is evaluated with Metropolis algorithm which
is known to suffer from critical slowing down. We also see by comparing
Figs. \ref{qcomp1} and \ref{qcomp2} that there is noticeable improvement 
of data with volume near the critical region.   

\section{Summary and conclusions}
In this work we have investigated the topological charge in 
1+1 dimensional lattice $ \phi^4$ field theory. We have shown that with
APBC in the spatial direction, lowest energy
configuration is a kink or an antikink. In order to characterize the
different phases of the theory with APBC we have proposed a simple
order parameter $\overline \phi_{\rm diff}$ and we have
demonstrated its effectiveness.  

In the process of computing the topological charge we have observed that 
as the system moves on from ordered to disordered phase, the single kink (or
antikink) gives way to the occurance of a multitude of kink-antikink pairs
suggesting kink condensation as a possible mechanism for this phase
transition.  

A major issue in extracting the topological charge is the renormalization.
By a particular choice of the renormalized coupling 
(in the topologically trivial sector), we are able to
express the topological charge in the renormalized theory as the ratio
of renormalized
order parameters in the lattice theories with APBC and PBC. Provided the
field renormalization constants are the same in the two cases, the
expression for the topological charge in the renormalized theory
becomes the ratio of
unrenormalized order parameters. Making use of this, we have computed the 
topological charge in the renormalized theory and demonstrated that it
indeed maintains
the value +1 in the broken phase. 

\acknowledgments
Numerical calculations presented in this work are carried 
out  on a Power4-based IBM cluster and a Cray XD1. The High
Performance Computing Facility is supported by 
the 10$^{th}$ Five Year Plan Projects of the Theory Division, Saha
Institute of Nuclear Physics, under the DAE, Govt. of India.  


\end{document}